\documentclass[review,floatfix]{elsarticle}

\usepackage{xspace}
\def\ie{i.e.\@\xspace}
\def\etc{etc.\@\xspace}

\usepackage{graphicx}

\journal{Nuclear Instruments and Methods in Physics Research Section A}

\begin{document}
\begin{frontmatter}

\title{Momentum and Energy Dependent Resolution Function of the ARCS Neutron Chopper Spectrometer at High Momentum Transfer: Comparing Simulation and Experiment}

\author[A]{S.O. Diallo\corref{correspondingauthor}}
\author[B]{J.Y.Y. Lin}
\author[C]{D.L. Abernathy}
\author[D,E]{R.T. Azuah}

\address[A]{Chemical and Engineering Materials Division, Oak Ridge National Laboratory, Oak Ridge, TN 37831}
\address[B]{Neutron Data Analysis and Visualization  Division, Oak Ridge National Laboratory, Oak Ridge, TN 37831}
\address[C]{Quantum Condensed Matter Division, Oak Ridge National Laboratory, Oak Ridge, TN 37831}
\address[D]{NIST Center for Neutron Research, Gaithersburg, Maryland 20742-2115, USA}
\address[E]{Department of Materials Science and Engineering, University of Maryland, College Park, Maryland 20742-2115, USA}
\cortext[correspondingauthor]{omardiallos@ornl.gov}

\begin{abstract}
 Inelastic neutron scattering at high  momentum transfers (i.e. $Q\ge20$ {\AA}), commonly known as deep inelastic neutron scattering (DINS), provides direct observation of the momentum distribution of light atoms, making it a powerful probe for studying single-particle motions in liquids and solids. The quantitative analysis of DINS data requires an accurate knowledge of  the instrument resolution function $R_{i}({Q},E)$ at each momentum $Q$ and energy transfer $E$, where the label $i$ indicates whether the resolution was experimentally observed $i={obs}$ or simulated $i=sim$.   Here, we  describe   two independent methods for determining the total resolution function $R_{i}({Q},E)$ of the ARCS neutron instrument at the Spallation Neutron Source, Oak Ridge National Laboratory. The first method uses experimental data from an archetypical   system  (liquid $^4$He) studied with DINS, which are then numerically  deconvoluted using its previously determined intrinsic scattering function to yield $R_{obs}({Q},E)$. The second approach uses accurate Monte Carlo simulations of the ARCS spectrometer, which account for all instrument contributions, coupled to a representative scattering kernel to reproduce the experimentally observed response $S({Q},E)$. Using  a delta function as scattering kernel, the simulation yields a resolution function $R_{sim}({Q},E)$ with comparable lineshape  and features as $R_{obs}({Q},E)$, but somewhat narrower due to the ideal nature of the model. Using each of these two $R_{i}({Q},E)$ separately, we extract characteristic parameters of liquid $^4$He such as the intrinsic linewidth $\alpha_2$ (which sets the atomic kinetic energy $\langle K\rangle\sim\alpha_2$) in the normal liquid and the Bose-Einstein condensate parameter $n_0$ in the superfluid phase.  The extracted $\alpha_2$ values agree   well with previous measurements at saturated vapor pressure (SVP) as well as at elevated pressure (24 bars) within experimental precision, independently of which $R_i(Q,y)$ is used to analyze the data.  The actual observed $n_0$ values at each $Q$ vary little with the model $R_{i}(Q,E)$, and  the effective $Q$-averaged $n_0$  values are consistent with each other, and with previously reported values.

 \end{abstract}


\end{frontmatter}

\section{Introduction}
Due to the unique properties of the neutrons,  their use as an experimental probe for studying atomic (or molecular) vibrations and interactions is a rather well-established technique which has contributed to many advances in various scientific areas such as condensed matter physics and chemical sciences \cite{Silver1989,Andreani2005}. Notable recent examples have revealed the existence of a magnetic resonance peak in iron-based superconductors \cite{Christianson2008},  clarified the connection between Bose-Einstein condensation and superfluidity in the Bosonic liquid $^4$He \cite{Diallo2014}, unravelled the existence of a roton-like signature in non-Bosonic liquid $^3$He \cite{Godfrin2012}, found a non-Gaussian proton momentum distribution in confined water \cite{Garbuio2007,Pietropaolo2006}, \etc.. 
The momentum ${Q}$ and energy $E$ landscape that can be probed by neutrons is rather broad, and generally not accessible by a single instrument and multiple reconfigurations of the same instrument \cite{Deen2015}are often necessary. This is because each neutron instrument is tailored and optimized (by design and technical limitations) to only probe a small cross-section of the wider ${Q}-E$ space, making it often necessary to  use multiple neutron instruments with overlapping $Q-E$ windows  and/or other complementary  techniques before a particular phenomenon can be fully understood. One notoriously known limiting factor for reconciling and interpreting neutron scattering data from different instruments is the fact that the  energy resolution function of a given instrument depends not only on the instrument parameters but also on the imparted momentum $Q$ and energy $E$ to the sample. Therefore, while qualitative interpretation of the raw INS data is generally possible (peak positions, dispersive nature of the excitations \etc), any rigorous quantitative analysis of INS data (excitation lifetimes, BEC fraction \etc) requires an accurate knowledge of the resolution function $R_{i}({Q},E)$ \cite{Andersen1996,Seeger2004,Vickery2013,Deen2015}.

\begin{figure}
 \includegraphics[width=0.9\linewidth]{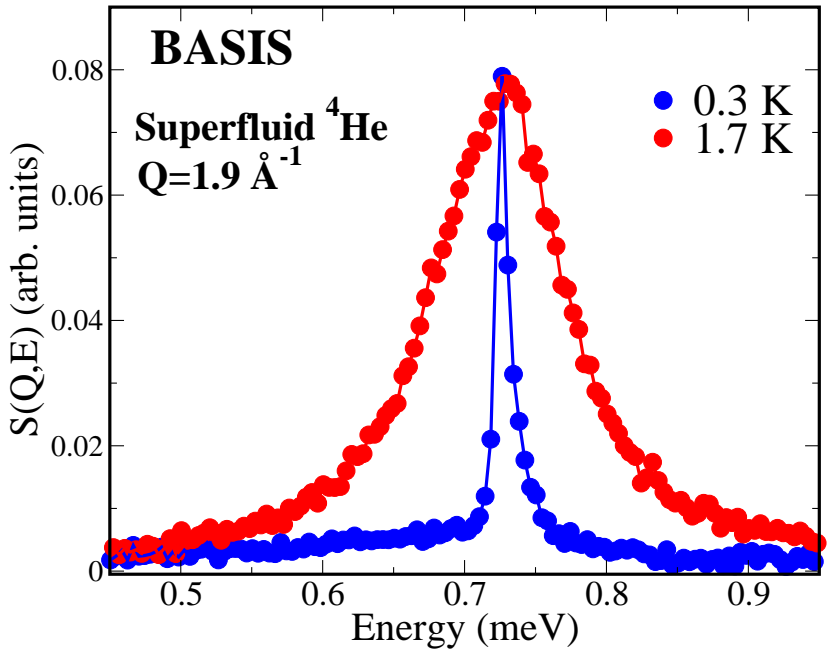}
\caption{Temperature dependence of the roton mode in superfluid $^4$He, as observed on the BASIS neutron spectrometer at the Spallation Neutron Source. The resolution limited asymmetric sharp peak (blue circles)  at low temperature is used to represent the intrinsic  energy resolution function (HWHM$\sim$ 5.8 $\mu$eV) for studying the temperature dependence of the roton lifetime at $T=1.7$ K (red circles).}
 \label{fig:Roton_BASIS}
 \end{figure}

To set the background, we note that at small  momentum transfers $Q$ (\ie $Q\le 2 $ \AA$^{-1}$), inelastic neutron scattering (INS) is quite effective in examining molecular re-orientation, diffusion processes and low energy excitations. In those cases, the resolution function is usually obtained by directly measuring the exact same sample at the lowest possible temperatures where  dynamical processes become frozen out on the instrument measurement time window, leaving out only the instrument contributions at or close to the elastic region. This is particularly true for backscattering instruments where the energy resolution width remains largely constant over the accessible $Q$ range and over a fairly broad dynamics range close to the elastic peak \cite{Tsapatsaris2015}. Fig. \ref{fig:Roton_BASIS} compares the observed elementary excitation (or \lq roton') in superfluid $^4$He at temperature of 1.7 K to its resolution limited response at 300 mK on the BASIS neutron spectrometer \cite{Mamontov2011} at the Oak Ridge National Laboratory's Spallation Neutron Source (SNS).  The low temperature measurement faithfully reproduces the asymmetric resolution due to the liquid H moderator source.  On most indirect geometry neutron instruments such as BASIS, or direct geometry  instruments in low $Q$ mode such as the DCS \cite{Copley2003} at the NIST center for neutron research, it may at times be sufficient to simply measure a purely incoherent standard such as vanadium in lieu of the low temperature measurements to determine $R_{i}({Q},E)$.  Unfortunately, no such straightforward practical approach exists when dealing with INS data from direct geometry instruments at high $Q$'s. At intermediate  $Q$'s (\ie $1\le Q\le 10$ \AA$^{-1}$), which is well-suited for studying  collective excitation modes such as phonons and magnons, and characteristic excitations such as rotons and molecular crystal fields, there exists few benchmarked software tools that account for resolution contributions in analyzing INS data collected on time-of-flight chopper spectrometers. The TOBYFIT software package \cite{Perring} which uses a semi-empirical method to approximate the resolution function has been the primary workhorse for analyzing time-of-flight INS data  in this regime. On the other end of the $Q$ spectrum, \ie the high momentum transfer regime ($Q\le 20 $ \AA$^{-1}$), which is the subject of this study, the inelastic scattering process of the individual  atoms resembles closely the scattering of freely moving particles and the neutron response is characterized by recoil scattering. In these two cases, $R_{i}({Q},E)$ can either be inferred from measuring calibrated samples for which the scattering function is well known so that it can be numerically deconvoluted from the measurements, or from using ray tracing Monte Carlo methods with all  instrument characteristics as input.  In this article, we present  the resolution function of the ARCS neutron spectrometer at the Spallation Neutron Source \cite{Abernathy2012} at the Oak Ridge National Laboratory obtained using each of these two procedures. The main goal is to properly account for the instrument resolution contributions when analyzing INS data obtained at high $Q$'s (i.e. DINS). Using these differently obtained $R_{i}({Q},E)$, we investigate the changes in the resulting average kinetic energy $\langle K\rangle$ of the atoms in normal liquid $^4$He, and the fraction of atoms that Bose condense (BEC) in the superfluid phase. While $\langle K\rangle$  is directly proportional to the linewidth $\alpha_2$ of the DINS signal, the macroscopic number of atoms in the BEC state can be inferred from the relative change in intensity in the DINS response between the normal and superfluid phases.  Our analysis  shows consistent $\langle K\rangle$  results  with previous measurements at saturated vapor pressure (SVP) \cite{Glyde2000} and at elevated pressure near the liquid-solid transition line \cite{Diallo2012}. The condensate fraction $n_0$ shows however sensitivity to otherwise marginally different $R_{i}({Q},E)$. 

\section{Neutron Measurements}
	\subsection{ARCS: The Wide-Angle Neutron Chopper Spectrometer}
\begin{figure}
\includegraphics[width=0.85\linewidth]{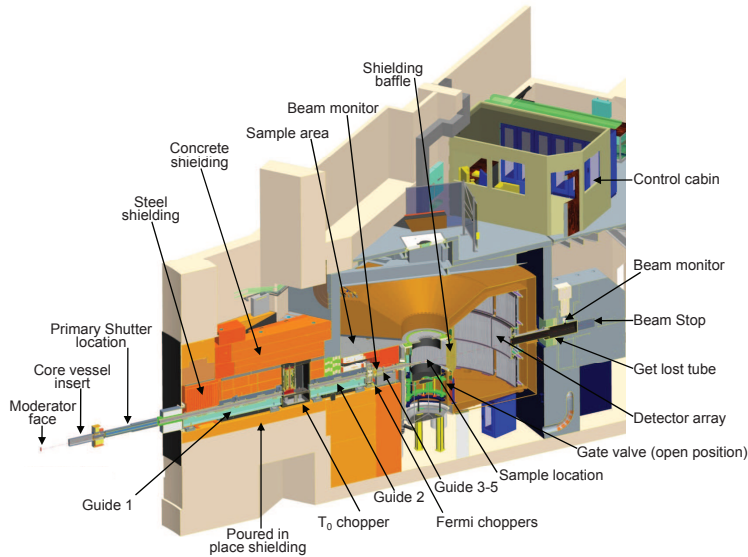}
\caption{Model layout of the ARCS wide-angle neutron chopper spectrometer at the Spallation Neutron Source (Oak Ridge National Laboratory).}
\label{fig:layout}
\end{figure}

	The wide angular range chopper spectrometer ARCS  \cite{Abernathy2012,Stone2014} is one of four direct geometry neutron instruments located at the Spallation Neutron Source (SNS), Oak Ridge National Laboratory. This means that it uses a monochromatic beam of neutrons whose energy $E_i$ can be set by the experimenter by varying the rotation speed and phase of a spinning Fermi chopper \cite{Voigt2014} located before the sample position to probe large area of momentum-energy ($Q-E$) space of a subject material. A model view of the ARCS instrument layout is illustrated in  Fig. \ref{fig:layout} for more details. The final energy $E_f$  of the neutrons after scattering from the sample is determined by time-of-flight techniques (TOF), allowing the energy transfer $E$ to the sample to be calculated, $E=E_{i}-E_{f}$. Using kinematic constraints for a given incident energy $E_i$, the momentum transfer $Q$ can be conveniently expressed as a function of  the energy transfer $E$, and the scattering angle $\phi$, \ie the angle between the incident and scattered beam (often called $2\theta$ in diffraction methods),  yielding,
	\begin{equation}
Q^2 = \frac{1}{\gamma} \left( 2E_i-E-2\sqrt{E_i(E_i-E)}cos\phi\right)
\label{eqn:QvsOmega}
\end{equation} 
\noindent where  $\gamma= \frac{\hbar^2}{2m_n}=$ 2.017 meV {\AA$^{2}$}.  On ARCS, this parametric relation  allows large  region of $Q-E$ to be probed, thanks to the large $\phi$ coverage by the detector arrays, $-28^\circ\le\phi\le135 ^\circ$. 

\begin{figure}
	 \includegraphics[width=0.8\linewidth,angle=90]{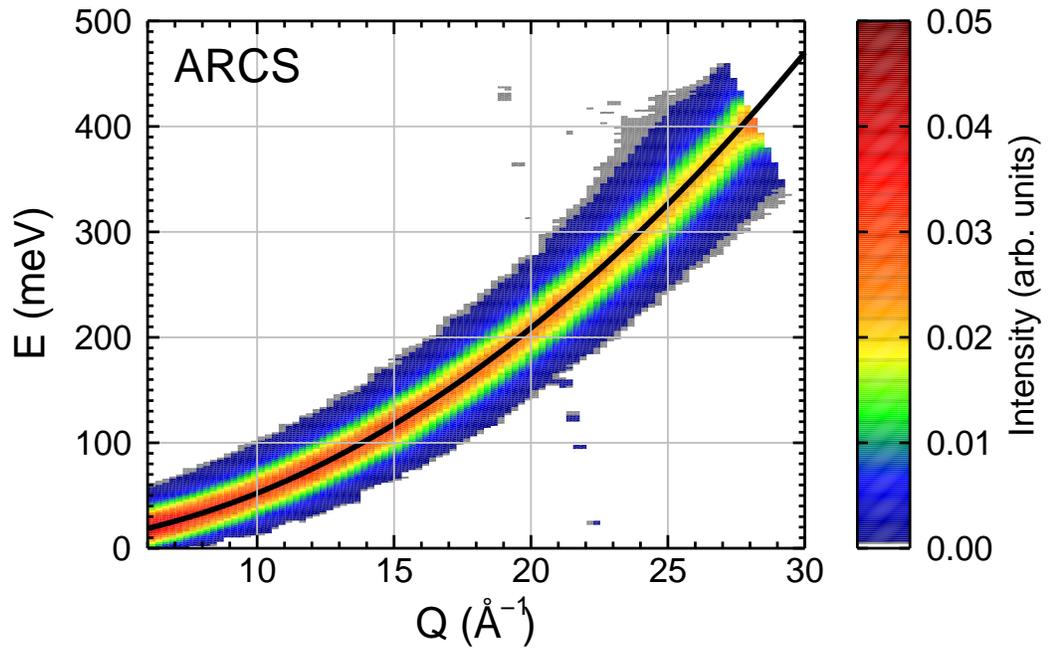} 
	   \caption{Background subtracted two-dimensional (2D) map of bulk normal liquid $^4$He under its own saturated vapor pressure (SVP) showing the observed instrument broadened $^4$He recoil signal (centered at $E_r=\frac{\hbar^2Q^2}{2m}$, shown as solid black line) in the energy-momentum ($E-Q$) space using an incident energy of $E_i\simeq$ 686 meV on the ARCS neutron chopper spectrometer.}
	 \label{fig:SQE_LHe_Exp}
	 \end{figure}

Fig. \ref{fig:SQE_LHe_Exp} illustrates the large $Q-E$ range accessible on ARCS  with the incident energy $E_i=$686 meV we used in our experiment. This particular two-dimensional (2D) map shows the data collected on bulk liquid $^4$He under its own vapor pressure (SVP) (colored curved strip) after proper background subtraction. Parts of the inaccessible regions due to kinematic constraints (parabola cut-off  imposed by Eq.\ref{eqn:QvsOmega}) are marked by the sharp edges on the strip at the high $Q$ and $E$ regions.  The solid black line is the ideal free $^4$He atom recoil line, expressed as $E_r=\hbar^2 Q^2/2m_{He}$.  The background considered here consists of 1) the empty Al container measured using the same instrumental configuration at 2.5 K and 2) an intrinsic weak broad angle-independent ($\Phi$) signal centered around 300 meV (less than 1-2\% of the overall intensity), suggestive of multiple scattering effects. This broad component was approximated by averaging out the total signal at low angles ($10\le \Phi\le30 ^\circ$, away from the $^4$He recoil scattering), and extrapolating it uniformly across all angles.  
Below we describe in details the neutron measurements and the MC simulations.

	 \subsection{High momentum transfer regime: Impulse Approximation}
	  In the high $Q$-regime also known as impulse approximation limit, the incoming  neutron transfers high energy and momentum to the atoms in the sample. In this event, the energy transferred to the sample, $E$, is large compared to the collective excitations energies in the material, with very short scattering time (atto-seconds) . The validity of this approximation in neutron scattering is very well-documented \cite{Silver1989,Andreani2005}.  The IA effectively treats the scattering event as single atom \lq billiard ball' scattering, in which the momentum and energy conservation rules apply to the neutron and target atom pair. As a result, the target atom with mass $m$ recoils somewhat independently of its neighbors with an energy $E_r=\frac{1}{2m}\hbar^2Q^2=\frac{m}{2} v_r^2$ (where $v_r=\frac{\hbar{Q}}{m}$ is the recoil speed) and the observed scattering intensity is well approximated by the incoherent dynamic structure factor $S(Q, E)$. In most liquids, this is an excellent approximation for Q$\ge$ 15 {\AA}$^{-1}$.  The scattering function $S(Q,E)$ can be expressed in terms of the longitudinal momentum variable, $y$, such that
\begin{equation}
J_{IA}(y)=v_rS({Q},E)
\label{eqn:jia}
\end{equation}
where $y = (E-E_{r})/\hbar v_{r}$ is essentially the component of the single particle atomic momentum projected along Q and $J_{IA}(y)$ is the longitudinal momentum distribution. This transformation reduces the separate $Q$ and $E$ dependence of the scattering into a single convenient variable $y$ and $J_{IA}(y)$ collapses into a $Q$-independent distribution that is centered around $y=0$. Unfortunately, the limits of the IA are not quite achieved experimentally and hence deviations known as final state effects (FSE) must be taken into considerations when analyzing the data. The FSE  account for the interactions of the struck $^4$He atom with its neighbors, as sensed by the neutrons upon scattering off the sample at these high but finite $Q$ values \cite{Glyde2000,Glyde2011a}. In $^4$He, the FSE do not change much with temperature and we use the previously determined FSE found at SVP  \cite{Glyde2000,Glyde2011} to treat the current DINS data. The FSE impose some residual $Q$ dependence on the observed scattering and so what is measured can be thought of as a convolution between $J_{IA}$ and FSE such that $J(Q,y) = J_{IA}(y){\otimes}FSE$. Hence careful analysis of $J(Q,y)$ over a sufficiently large range in $Q$ can yield information on both $J_{IA}(y)$ and FSE since J$_{IA}$ is $Q$-independent and FSE are not.
 For $^4$He with $m_{He}=$4.002 gram/mol, we use the  factor $\lambda_{He}=\hbar^2/m_{He}=$1.0443 meV {\AA}$^2$ to convert $\sigma_E$ to $\sigma_y$. Similarly, for protonated compounds (e.g. H$_2$O) commonly studied in DINS, it is convenient to use the conversion factor $\lambda_{H}=4\times\lambda_{He}=4.17$ meV {\AA}$^{2}$ since $m_{H}=\frac{1}{4}m_{He}$= 1 gram/mol. As a side note, it is worthwhile noting that while H$_2$O related compounds are measurable on ARCS using 2-3 eV incident energy, the H recoil line is significantly broader than the instrument resolution and not limited by resolution effects. However, because of the relatively smaller mass of hydrogen compared to $^4$He, the recoil line doesn't extend high enough in $Q$ and the reliable DINS analysis tends to be limited to very small $Q$ range.  To study single proton dynamics over a wider high $Q$-range on neutron chopper instruments  such as ARCS or the sister instrument SEQUOIA\cite{Senesi2014}, it may be useful to use deuterated compounds (D) to shift the proton recoil line to lower energies and access a wider $Q$ range. 

In general, the actual observed scattering function $J_{obs}(Q,y)$ is the resolution broadened signal $J(Q,y)$,
\begin{eqnarray}
J_{obs}(Q,y)&=&J_(Q,y)\otimes R_i(Q,y)  \nonumber \\
		&=&\int J(Q,y')R_i(Q,y-y')dy'  
\label{eqn:jqy_conv}
\end{eqnarray}
\noindent Eq. \ref{eqn:jqy_conv} can be computed numerically if  $J_(Q,y)$ and $R_i(Q,y)$ are known. In practice however, only $J_{obs}(Q,y)$ and $R_i(Q,y)$ are directly accessible.  This means that $J(Q,y)$ can only be obtained by inverting Eq. \ref{eqn:jqy_conv}   using deconvolution methods.  Cumulant expansion methods\cite{Silver1989,Reiter2002,Glyde1994,Watson1996,Senesi2012} are generally used to obtain analytical expressions for $J(Q,y)$, often up to $6^{th}$ order depending on the need and the data quality.     As explained above, in cases where a measured $R_i(Q,y)$ is not available, it becomes necessary to use a simulated $R_i(Q,y)$ to extract $J_{IA}(y)$ and FSE.  The intrinsic $J_{IA}(y)$ for bulk liquid $^4$He is well documented with a well known temperature dependent linewidth $\sigma_y(Q)$ \cite{Glyde2000,Andreani2005}.   At $Q=20$ \AA$^{-1}$ for example, the intrinsic $^4$He linewidth  in energy ($1^{st}$ leading Gaussian term in the cumulant expansion) is $\sigma_{E}$=$20\times1.0443\times\sigma_y$=$20.88\times0.9$= 18.8 meV at SVP. 
In the present analysis, we retain the full 6$^{th}$ order expansion of $J_{IA}(y)$ and FSE, and allow only the 1$^{st}$ leading term ($\alpha_2$) and condensate component to vary. The higher order terms in $J_{IA}(y)$ ($\alpha_4$ and $\alpha_6$) are kept fixed at their previously determined values at SVP and at 24 bars \cite{Glyde2011}.

	 \subsection{Observations in liquid $^{4}$He}
We used ARCS in its high energy resolution configuration with $E_i=686$ meV to gather DINS data of normal liquid state ($T=2.5$ K) at SVP as a benchmark measurement. Without actual superfluid data at SVP from ARCS, we use instead our previously collected superfluid data at elevated pressure of 24 bars \cite{Diallo2014} to determine the condensate fraction.  The data analysis was subsequently performed over the $Q$-range, $21.5\le Q\le27.5$ {\AA}$^{-1}$ with $\delta Q=$0.5 {\AA}$^{-1}$, which we found to have the best signal-to-background ratio in the present measurement. 
  	
	\begin{figure}
\includegraphics[width=0.9\linewidth]{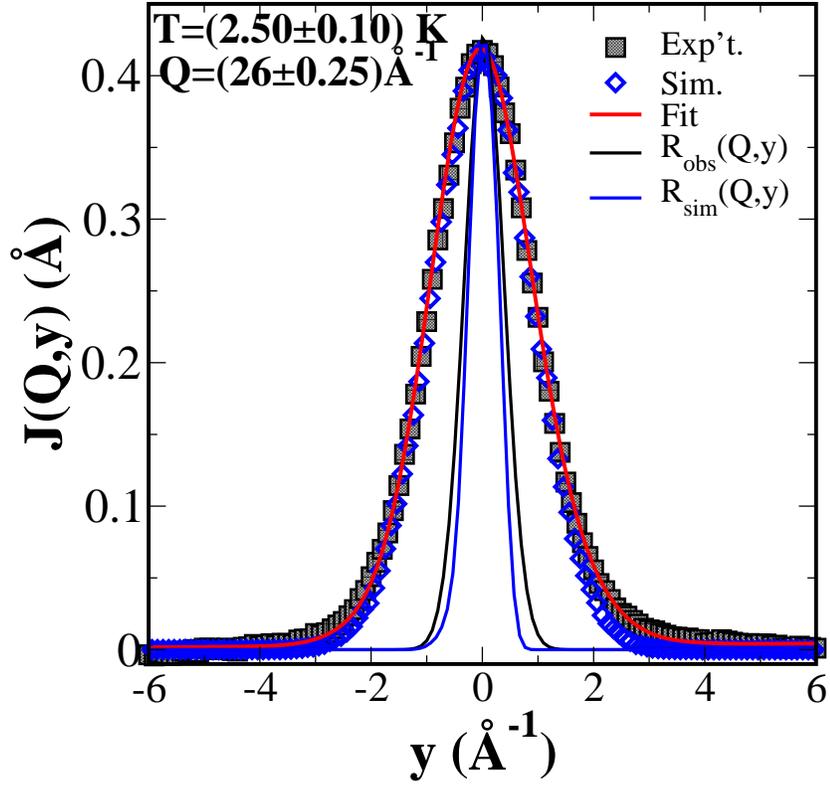}
	  \caption{Exemplary 1D cut taken along  the scale variable $y\propto (E-E_r)$ at $Q=(26\pm0.25)$ {\AA}$^{-1}$ comparing the actual measurement (filled square symbols) with the simulated response (open blue diamonds) and the different resolution functions $R_{i}(Q,y)$. The blue solid line ($R_{sim}(Q,y)$) represents the simulated resolution function and the black solid line ($R_{obs}(Q,y)$) is the one derived from experiment. The red solid line is the resolution convoluted fit, using $R_{obs}(Q,y)$. Similar fit is obtained with $R_{sim}(Q,y)$.}
	 \label{fig:SQE_cuts}
	 \end{figure}
	 
	 Fig. \ref{fig:SQE_cuts} shows a representative one-dimensional (1D) spectra cut  taken around $Q=(26\pm0.25)$ {\AA}$^{-1}$ as a function of the longitudinal momentum  variable $y\propto (E-E_r)$. This  figure compares the net experimental scattering intensity of liquid $^4$He with the simulated response using the MCViNe ray tracing Monte Carlo, described in Ref.\cite{Lin2014,Lin2016} to which we return below. The red line is the best model fit (using Eq. \ref{eqn:jqy_conv}) with $R_i(Q,y)=R_{obs}(Q,y)$. A very similar fit can be also obtained  with $R_{sim}(Q,y)$, but with slightly different fit parameters.  To truly appreciate the difference in the quality of the fits and for better comparison, we undertook a detailed $\chi^2$ analysis as a function of $Q$, and could not again satisfactorily differentiate between the fitting agreement factors $\chi^2$, other than to say that  both $R_{i}$ yield consistent fits at most $Q$ values, as indicated in Table \ref{tbl:chisquared} for superfluid $^4$He at 24 bars.

 \begin{table}
\caption{Fitting agreement factor $\chi^2(Q)=\sum_l \frac{[J_{{obs}_l}(Q,y)-J_{{model}_l}(Q,y)]^2}{n_p-m}$ for superfluid $^4$He data at 24 bars versus selected $Q$ for different resolution function $R_{i}$, where the label $i$ refers to the extraction method used to determine $R_i(Q,y)$. The parameter $n_p$ represents the number of data points and $m$ the number of fitting parameters in $J_{model}(Q,y)$.}
\begin{center}
\begin{tabular}{| c | c | c |  c | c |c | c |}
\hline
\hline
$Q$ (\AA$^{-1}$) & 22.0  &23.0  & 24.0  & 25.0  & 26.0  &27  \\
\hline
$R_{obs}$  &1.05  & 0.88 &1.10 &1.09 &2.95& 1.9\\
$R_{sim}$ & 0.93  & 0.81 & 1.01 & 1.06 & 2.99 & 2.02\\
\hline
\hline
\end{tabular}
\end{center}
\label{tbl:chisquared}
\end{table}	 


	 \subsection{Measured resolution function}
\begin{figure}
 \includegraphics[width=0.9\linewidth]{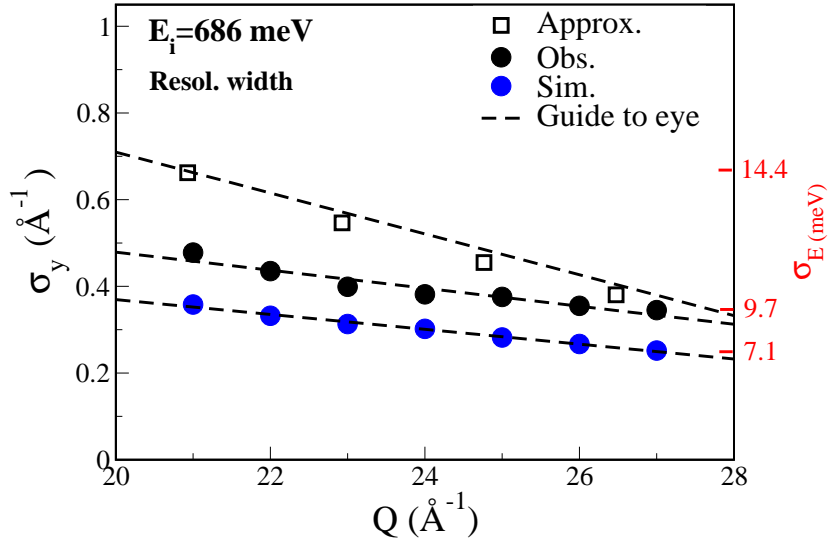}
\caption{$Q$-dependence of the Gaussian linewidth $\sigma_y$ of the ARCS resolution function. The black circles are the experimentally extracted widths, while the blue circles are the fitted Gaussian widths of the MCViNE simulations. The open symbols are approximated values for the expected line broadening along the recoil line, based on source, and instrument component considerations (see Eq. 1 in Ref. \cite{Abernathy2012}). As a visual guide, we convert $\sigma_y$ to $\sigma_E$ using $\sigma_E=1.0443\times Q \times\sigma_y$, which gives the unevenly spaced vertical scale on the right. In all cases, the resolution linewidth decreases with increasing $Q$, as expected.}
\label{fig:Res_Exp_vsSim}
 \end{figure}

	To  extract the ARCS resolution function from measurements, we use the known intrinsic $J_{IA}(Q,y)$ of normal liquid $^4$He at SVP \cite{Andreani2005,Glyde2000} convoluted with a parametric resolution function to reproduce the observed resolution-broadened $J(Q,y)$ of normal $^4$He by least-square fitting methods. The resolution function was parametrized with an empirical function consisting of  up to no more than two Gaussians. We begin our fits assuming a single Gaussian model $R_i(Q,y)$ first.  This simple model does reproduce our data fairly well, as evidenced by the resulting $\chi^2$ values, and the fit lines. In an attempt to further these already excellent fits, we add a second Gaussian component but keep the first one fixed. With this constraint however, we found the second Gaussian component to be largely insignificant and roughly an order of magnitude weaker. For this reason, we found little to no difference in the overall energy resolution width with the addition of this second Gaussian, and use only the single Gaussian model for the rest of the analysis. The net resolution width, denoted $\sigma_y$ in the $y$-space, generally decreases with $Q$ and is narrower than the $^4$He signal over the entire $Q$ range spanned in the present experiment. This observed $\sigma_y$ on ARCS is depicted in Fig. \ref{fig:Res_Exp_vsSim} for selected  $Q$ values, along with the simulated resolution value obtained using the MCViNE software package (described below and in Ref.\cite{Lin2016}). While the simulated $\sigma_y$ is consistently smaller than the measured one (1-2  meV sharper in energy), again most likely due to the ideal nature of our scattering kernel, the overall behavior with $Q$ remains the same; essentially, $\sigma_y$ decreases with increasing $Q$, making the resolution finer at the high energies along the recoil line. Specific details regarding the simulations are provided in Section \ref{sec:sim}. We are thus interested in assessing the impact of such a systematic difference in $R_i(Q,y)$ on the quantitative outcomes of subsequent data modeling.  Our chief aim here  is to evaluate the characteristic parameters of liquid $^4$He (such as the kinetic energy and the condensate fraction) using these resolution functions, and determine the most reliable approach  for analyzing DINS data on the ARCS instrument. To better put these resolution widths in context, we note for example that the intrinsic Gaussian line broadening of normal liquid $^4$He is $\sim$24.4 meV at $Q=$27 \AA$^{-1}$, while the measured and simulated resolution width $\sigma_y$ are respectively 9.7 and 7.1 meV. Adding these widths in quadrature leads to consistency  with the observed width for the $J(Q,y)$ shown in Fig. \ref{fig:SQE_cuts}.

\section{Monte Carlo Ray-tracing Simulations \label{sec:sim}}

The MCViNE software package \cite{Lin2016} was used to perform the Monte Carlo Ray-tracing simulations of the neutron scattering experiments, as illustrated for example in Fig. \ref{fig:SQE_cuts}.  In this section, we discuss the simulation procedure, and main simulation components that are critical in the simulations.

	\subsection{Simulation procedure}
The MCViNE software package \cite{Lin2016} is an object oriented (OO) Monte Carlo neutron ray tracing package for modeling and simulation of neutron scattering experiments on modern neutron spectrometers as described in detail in Refs. \cite{Lin2014,Yiu2016,Abernathy2012}. MCViNE simulations were performed in four steps to reproduce the experiments in high fidelity:

\begin{enumerate}
\item Beam simulation: The incident beam on the sample was simulated using the same benchmarked instrument model used in previous research \cite{Lin2014,Yiu2016,Abernathy2012}.
\item Sample scattering: The neutron packets simulated in the previous step were sent to the $^4$He sample simulation component (details below),  and the scattered neutrons were saved.
\item Detector interception: Each scatter neutron is sent to a virtual detector system consisting of three arrays of detector banks (each array consists of many bundles of eight $^3$He tubes known as 8-packs) located at the same positions and having the same orientations as the actual physical detector system.  The scattered neutrons are recorded by the time-of-flight bin indexes and pixel ID values, and event-mode NeXus data files are generated.
\item Reduction: The simulated NeXus data is reduced by Mantid \cite{Arnold2014} following the same procedure as the experiment data for consistency.
\end{enumerate}

\subsection{Neutron beam simulation components}
\begin{figure}
 \includegraphics[width=1\linewidth]{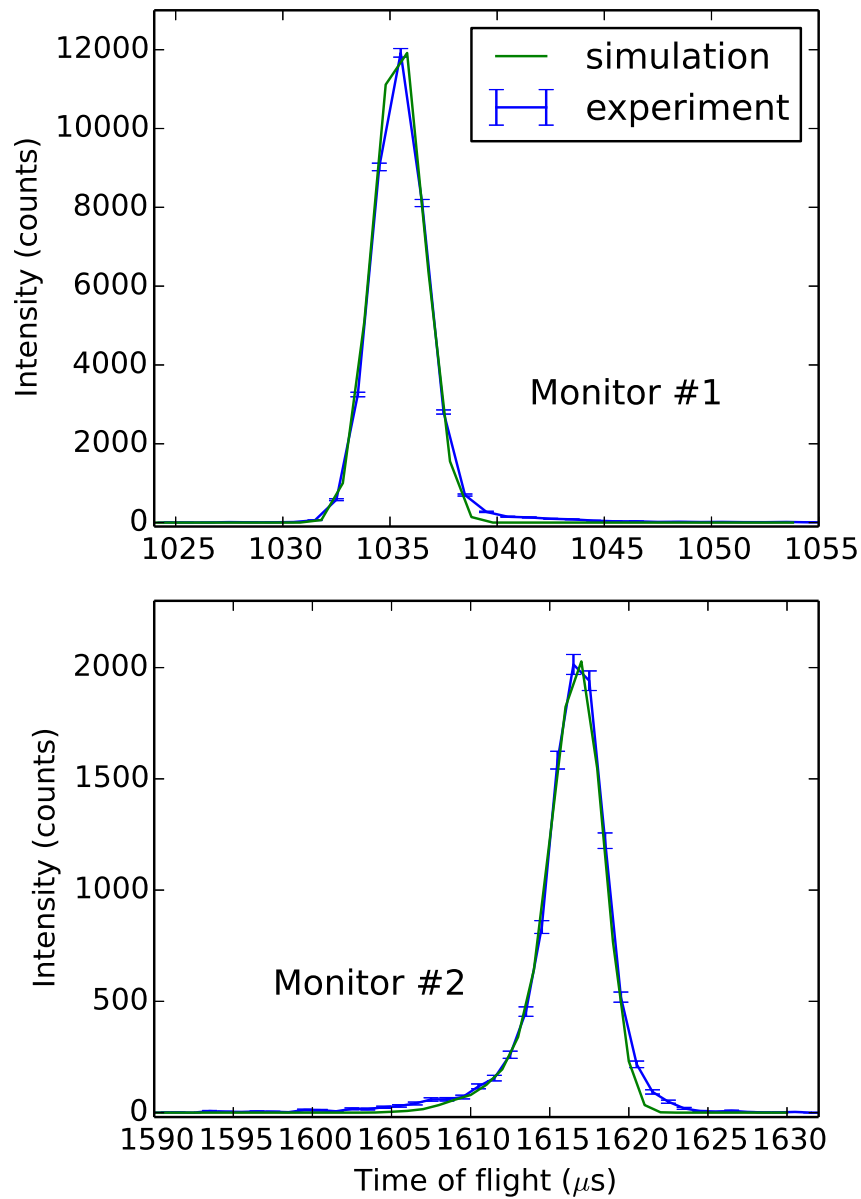}
\caption{Beam monitor peaks (blue lines). {\bf Upper:} Spectra from monitor 1, located between the Fermi Chopper and the sample (see Fig. \ref{fig:layout}). This monitor captures the beam profile before the neutrons reach the sample. {\bf Lower:} Spectra from monitor 2, located between the sample and the detectors. This second monitor best captures the moderator speed-time distribution profile, from which the IKeda-Carpenter (IC) parameters can be extracted.  The green lines show the simulated peaks using the IC parameters given in Table \ref{tbl:IC_parms}.}
\label{fig:monitor}
 \end{figure}

The neutron beam simulation starts with the moderator, continues with a series of neutron optics such as neutron guides and Fermi and T$_0$ choppers, and ends right before the sample position.
All components contribute to the shaping and broadening of the incident beam, but two components turn out to be most important: the moderator and the Fermi chopper. The moderator characteristics are modeled using parameterized Ikeda-Carpenter \cite{Loong1987, Abernathy2015} speed and time distribution functions [which can be found at \cite{McStas_ST}], which reproduced well the shape of the peaks in the beam monitors located before and after the sample (see Fig. \ref{fig:layout} for actual positions), as indicated in Fig. \ref{fig:monitor}. The Fermi chopper was simulated using the McStas \cite{Willendrup2004} Fermi chopper component \cite{McStas_FC} with the nominal parameters of the 0.5 mm slit spacing, 1.5 meter radius of curvature Fermi chopper used in the experiment. Small deviation of the Fermi chopper parameters in reality from the nominal values, and/or the slight numerical errors in the Fermi chopper components, may have contributed to the small deviation of the simulated resolution widths from the experimental values.
 
  \begin{table}
\caption{Parameters of the Ikeda-Carpenter moderator \cite{Ikeda1985} speed-time distribution used in the simulations for an incident energy $E_i=$686 meV. The parameters $\alpha$ and $\beta$ are the inverse time constants (in $\mu$s)  of fast and slow processes respectively in the neutron moderator and reflector system  \cite{Ikeda1985,Abernathy2015}. The variable  $R$ represents the fractional contribution of the slow process.}
\begin{center}
\begin{tabular}{| c | c | c |  c | c |c | }
\hline
\hline
$E_i$ (meV) & $\alpha$ ($\mu$s$^{-1}$)& $\beta$ ($\mu$s$^{-1}$) & $R$ \\
\hline
686 &  1.454047  & 0.213818 & 0.270462\\
\hline
\end{tabular}
\end{center}
\label{tbl:IC_parms}
\end{table}

\subsection{Sample simulation component}
The sample simulations accounted for the actual sample characteristics, namely the cylindrical shape (confined by the sample cell) with diameter of 2.5 cm radius and 8.55 cm height, and the SVP density and neutron cross section of $^4$He. The scattering kernel was assumed to be an isotropic dispersion function. Depending on the study, the kernel may contain an intrinsic broadening in the form of $1.0443\times0.89\times Q$ \cite{Glyde2000} (where $0.89$ refers to the intrinsic $y$-broadening of liquid $^4$He at SVP) for direct simulation of the experimental data, or may be an ideal $\delta$-function without any broadening for simulation of the resolution function itself.

The observed intrinsic resolution widths $\sigma_y$ are compared in Fig. \ref{fig:Res_Exp_vsSim}. The directly simulated $R_{sim}(Q,y)$ is slightly narrower than the measured $R_{obs}(Q,y)$,  in agreement with the widths observed from deconvoluting the full simulation of $^4$He (as done for the experimental data).  We also show the expected line broadening along the recoil line using   Ref. \cite{Abernathy2012}. These values are obtained by considering the contributions to the timing uncertainty from the source, chopper opening and path length differences to be statistically independent, which are then added in quadrature \cite{Abernathy2012}. This comparison confirms that the ARCS resolution improves with increasing $Q$.

	 \begin{figure}
	 \includegraphics[width=0.9\linewidth]{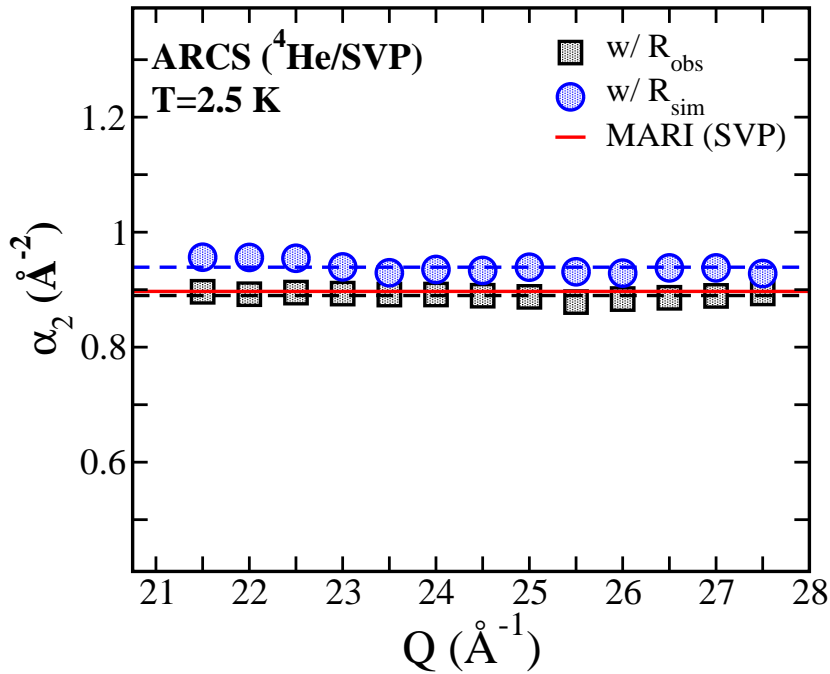}
	 \caption{Observed wavevector dependence of the characteristic width parameter $\alpha_2$  of liquid helium at temperature $T=2.5$ K and its own saturated vapor pressure (SVP). This parameter is directly proportional to the single particle kinetic energy ($\langle K\rangle=\frac{3}{2}\lambda\alpha_2$ where $\lambda=\hbar^2/m=$12.12 K {\AA}$^{2}$). Symbols are experimentally determined parameters and the lines correspond to their $Q$-averaged values. The dashed lines represent average values and the red solid line is that of bulk liquid $^4$He observed at SVP on the MARI spectrometer \cite{Glyde2000}.}
	 \label{fig:a2_SVP}
	 \end{figure}
	 
\section{Discussion}

\begin{figure}
	 \includegraphics[width=0.9\linewidth]{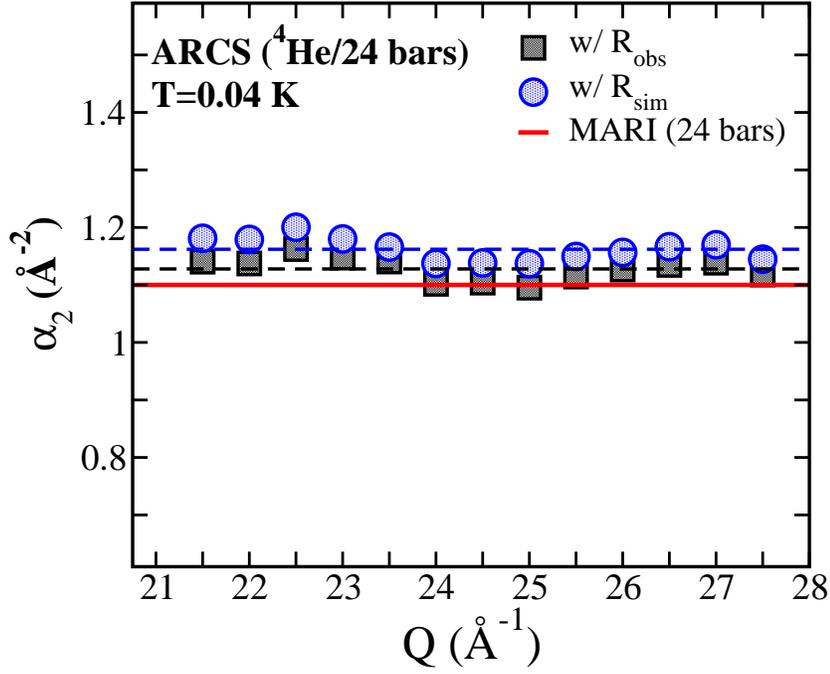}
	 \caption{Effect of instrument resolution on the wavevector dependence of $\alpha_2$  of liquid $^4$He at temperature $T=2.5$ K and pressure $P=24$ bars. The elevated pressure leads to an increase in density, which in turn leads to larger kinetic energy and $\alpha_2$ values compared to the bulk liquid values. The lines represent the $Q$-averaged values. The red solid line shows the average bulk liquid $^4$He value observed on MARI at 24 bars  \cite{Glyde2000}.}
	 \label{fig:a2_HP}
	 \end{figure}

	More generally, the resulting MCViNE simulated resolutions are a little sharper than the experimentally determined ones (see Fig. \ref{fig:SQE_cuts}), albeit  the excellent fits one can obtain with either. This  results in an absolute value of the linewidth $\alpha_2$ of liquid $^4$He that is consequently smaller with $R_{obs}$ than with $R_{sim}$. However, accounting for the standard errors, we conclude that these $\alpha_2$ values are in fact consistent with each other and with previous measurement \cite{Glyde2000}. Fig. \ref{fig:a2_SVP} shows the observed dependence on $Q$ of $\alpha_2$ in normal liquid $^4$He at $T=2.5$ K and at saturated vapor pressure (SVP). This fundamentally important variable is a good measure for the average atomic kinetic energy; $\langle K\rangle=\frac{3}{2}\lambda\alpha_2$ where $\lambda=\hbar^2/m$=12.12 K {\AA}$^{2}$.  
	Having verified that the method works well for liquid $^4$He at SVP, we then use it to understand the dynamical properties of $^4$He under pressure, a clean thermodynamic tuning variable. When liquid $^4$He  is subjected to elevated pressures, the resulting spatial localization of atoms ($\Delta x$ decrease) leads to an uncertainty in their position in momentum space ($\Delta p$ increase) due to quantum mechanics principles. Therefore, the atomic kinetic energy  ($\propto \alpha_2$) would be expected to increase as the pressure is raised towards the liquid-solid line, while the condensate fraction gets reduced due to the increased atomic interactions.    It is fundamentally interesting to understand how pressure affects  the local environment of individual $^4$He atoms, and to experimentally evaluate any pressure-induced changes in the average kinetic energy and the Bose-condensate parameter \cite{Diallo2012}.  Fig. \ref{fig:a2_HP} shows the extracted $\alpha_2$ value using liquid $^4$He data at 24 bars, analyzed using the two determined $R_i(Q,y)$.  The solid red line indicates again the reference value previously measured on MARI  (another direct geometry neutron spectrometer at the Rutherford Appleton Laboratory in the United Kingdom) at 24 bars.  While the observed $\alpha_2$ parameters are all expectedly larger than the bulk value, they compare favorably with the observations on MARI \cite{Glyde2011,Glyde2011a,Diallo2012} within experimental precision.  

\begin{figure}
	 \includegraphics[width=0.9\linewidth]{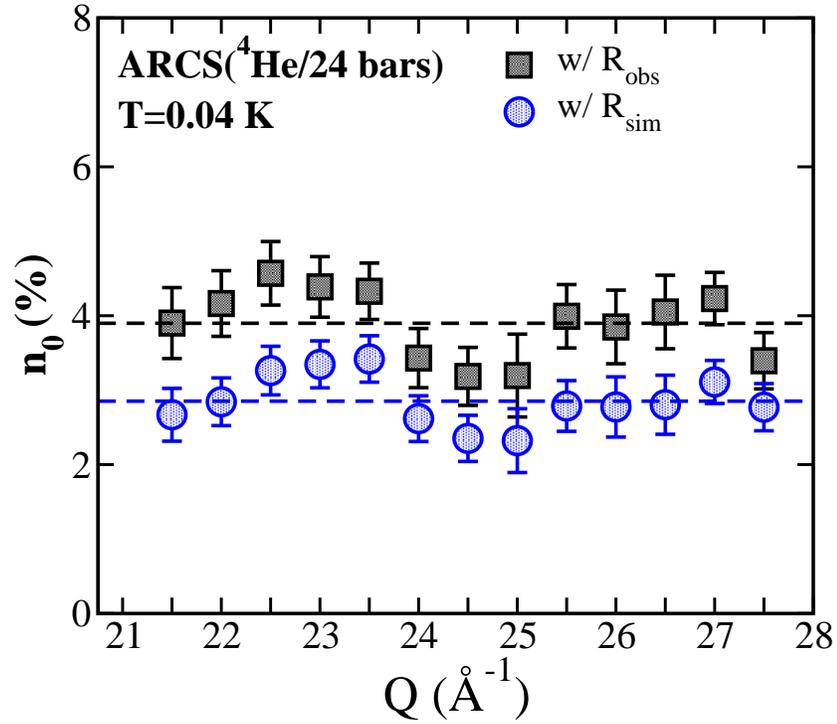}
	 \caption{Observed $Q$-dependence of the Bose condensate fraction $n_0$   of liquid helium at elevated pressure of $P=24$ bars and temperature $T=0.04$ K, based on the experimentally extracted (black solid squares and blue open diamonds) and calculated (red open circles) resolution functions. The lines represent the corresponding $Q$-averaged values. The variation of $n_0$ with $Q$ observed with $R_{obs}(Q,y)$ are also observed with $R_{sim}(Q,E)$. The dashed lines represent the average values.}
	 \label{fig:n0_HP}
	 \end{figure}

\begin{table}
\caption{Resolution dependence of the $Q$-averaged characteristic parameters (\ie linewidth $\alpha_2$) of $^4$He signal  and condensate fraction $n_0$) of superfluid $^4$He (0.04 K), as observed on the ARCS neutron time-of-flight chopper spectrometer. Results from MARI instrument are shown for comparison.}
\begin{center}
\begin{tabular}{| c | c |  c | c |}
\hline
\hline
T (K)		&	2.5	&	0.04	& 0.04	\\
\hline
Inst. Resol.    & $\alpha_2$ (SVP) &  $\alpha_2$ (24 bars) & $n0$ (\%) (24 bars) \\ 
\hline
ARCS $R_{obs.}$  & 0.88$\pm$0.02 & 1.13$\pm$0.02 & 3.89$\pm$0.47 \\
\hline
ARCS $R_{sim.}$ &  0.93$\pm$0.02 & 1.16$\pm$0.02 & 2.85$\pm$0.35 \\
\hline
MARI  $R_{sim}$  &  0.89$\pm$0.02  & 1.10$\pm$0.02 & 3.20$\pm$0.75 \\
\hline
\hline
\end{tabular}
\end{center}
\label{tbl:sum}
\end{table}	
	Finally, we observe a Bose-Einstein condensate fraction $n_0$ at 24 bars that is indeed smaller than the reported bulk value of 7.25\% at $T=0$ K \cite{Glyde2000}, independently of the  resolution function used to analyze the data. The observed $n_0$ values are shown as a function of $Q$ in Fig. \ref{fig:n0_HP}.  While the $n_0$ values derived from $R_{sim}(Q,y)$ tend to be generally smaller than those obtained with $R_{obs}(Q,y)$ in magnitude, the effective $Q-$averaged values are consistent with each other and with previous work \cite{Diallo2012}, as summarized in Table \ref{tbl:sum}.  The relative sensitivity of $n_0$ to resolution effects highlights the importance of reliably and accurately determining the full $R_{i}(Q,y)$ to better analyze DINS data. In the present study, either$R_i(Q,y)$ provides reliable characteristic parameters for the momentum distribution of liquid $^4$He which agree with theoretical predictions\cite{Ceperley1986,Ceperley1995,Moroni2004}, and  previous measurements \cite{Sokol1993,Glyde2000,Diallo2012}.

\section{Conclusion}
In this comparative study, we have presented  different experimental and simulation methods for determining the full resolution function of a direct-geometry time-of-flight spectrometer for the purpose of analyzing DINS data.   Independently of the methods used, the observed resolution broadened dynamical scattering patterns $J({Q},y)$ are consistently similar for liquid $^4$He. Consequently, the extracted signal width $\alpha_2$ values are consistent with each other and agree  well with previous work.  The average BEC values ($n_0$) at 24 bars are more sensitive to changes in the resolution lineshape than $\alpha_2$, varying from for example from $\sim$2.85\% with $R_{sim}(Q,E)$ to $\sim$3.9\% with $R_{obs}(Q,E)$ within acceptable uncertainty.  The simulations offer the most direct route for analyzing future DINS data at ARCS.  While the methods  presented are for analyzing DINS data for the ARCS spectrometer with liquid $^4$He as the sample, they can be effectively applied to other systems such as H$_2$ and H$_2$O measured either at ARCS or other neutron spectrometers such as SEQUOIA.

\section{Acknowledgments}
We wish to thank R. Senesi,  T. Prisk, E. Mamontov, H. Bordallo, F.X. Gallmeier, E. Iverson, G.E. Granroth,  B. Fultz and H. Glyde for many valuable stimulating discussions. We acknowledge the use of the Mantid software package \cite{Arnold2014} to reduce the neutron data and the NIST DAVE fitting software \cite{Azuah2009} for the data analysis.  This work is sponsored by the Scientific User Facilities Division, Office of Basic Energy Sciences, U.S. Department of Energy.

\end{document}